\documentclass[%
 reprint,
amsmath, amssymb,
aps,
prl,
]{revtex4-1}

\usepackage{graphicx}
\usepackage{dcolumn}
\usepackage{bm}
\usepackage{float}
\usepackage{epstopdf}

\begin{document}

\preprint{APS/123-QED}

\title{A Cascading Failure Model by Quantifying Interactions}

\author{Junjian Qi}
\author{Shengwei Mei}%
\affiliation{%
 Department of Electrical Engineering, Tsinghua University, Beijing, China 100084 
}%





\begin{abstract}
Cascading failures triggered by trivial initial events
are encountered in many complex systems.
It is the interaction and coupling
between components of the system that causes cascading failures.
We propose a simple model to simulate cascading failure
by using the matrix that determines 
how components interact with each other.
A careful comparison is made between the original cascades
and the simulated cascades by the proposed model.
It is seen that the model can capture general features of the original cascades,
suggesting that the interaction matrix
can well reflect the relationship between components.
An index is also defined to identify important links
and the distribution follows an obvious power law.
By eliminating a small number of most important links
the risk of cascading failures can be significantly mitigated,
which is dramatically different from
getting rid of the same number of links randomly.

\end{abstract}

\pacs{Valid PACS appear here}
\maketitle


Cascading failures in complex systems are 
complicated sequences of dependent outages. 
They can take place in electric power systems \cite{nerc, us blackout},
the Internet \cite{Internet1}, the road system \cite{road1}, 
and the social and economic systems \cite{social}.
For example, in power grid a line is tripped for some reason,
such as error of operators, bad weather, or tree contact, 
and the power transmitted through this line will be redistributed to other lines, 
possibly causing other line tripping or even cascading failures.

Several models have been proposed to study the general
mechanisms of cascading failures. 
The topological threshold models \cite{threshold} 
provide insight into several types of cascading failures.
The influence model \cite{influence} 
comprises sites connected by a network structure and 
each site has a status evolving according to Markov chain. 
In power systems OPA (ORNL-PSerc-Alaska) model \cite{OPA1, OPA2} and its variants, such as 
AC OPA \cite{AC OPA1, AC OPA2} and OPA with slow process \citep{slow1,slow2}, are proposed to
study the complex global dynamics of blackouts.
More recently the line interaction graph \cite{line graph} 
initiates a novel analysis method for cascading failures 
in electric power systems by considering the interaction of the transmission lines.

The system-level failure of tightly coupled complex systems 
is not caused by any specific reason 
but the property that components of the system 
are tightly coupled and dependent \cite{normalAccident}.
Based on this idea we quantify the interaction of the components 
by an interaction matrix and then 
propose a cascading failure model by using this matrix. 
The interaction matrix can be obtained with 
cascades from simulations or real statistical data.
These cascades can be grouped into different generations \cite{bp1, bp2, bp3} 
and are called original cascades to 
distinguish with the simulated cascades in the following discussion.
A typical cascade can be:

\begin{table}[H]
\renewcommand{\arraystretch}{1.8}
\label{data}
\centering
\begin{tabular}{cc}
generation\,0 & generation\,1 \\
25, 40, 74, 102, 155 & 72, 73, 82 \\
\end{tabular}
\end{table}

Here the numbers are the serial number of 
the failed components. This cascade has two generations 
while others might contain one or several generations.

Topological properties such as small-world \cite{small world} 
and scale-free \cite{scale free} behavior 
have been found in complex networks. 
Power-law behavior is also discovered 
for weighted networks \cite{weighted}. 
In this paper we discuss the property of 
the directed weighted interaction network rather than 
the network directly from the physical system.

\begin{figure}
\centering
\includegraphics[width=2.0in]{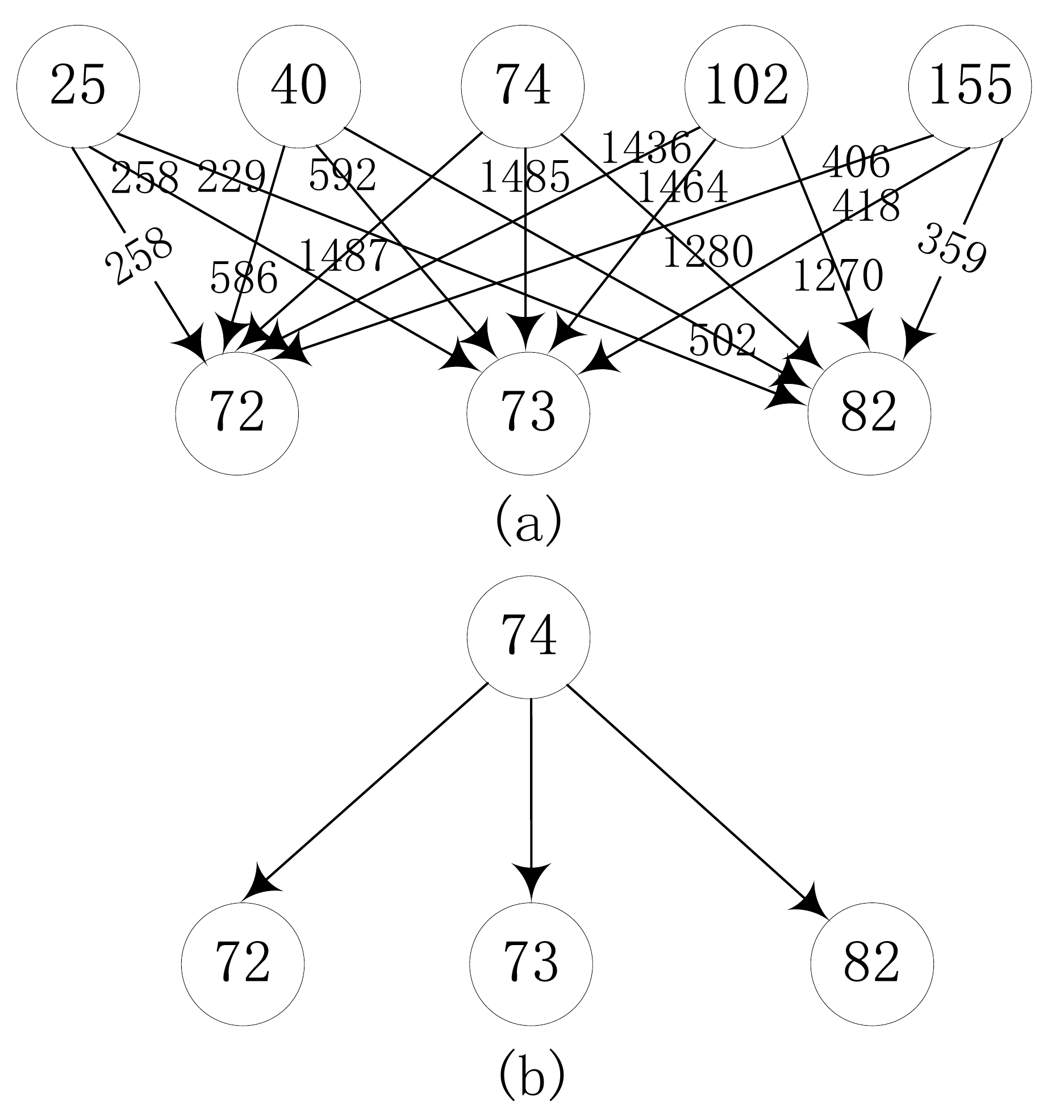}
\caption{Illustration of correcting $\mathbf{A}$ to $\mathbf{A}'$. Two consecutive generations of a cascade are given.
The last generation comprises the failure of 25, 40, 74, 102, and 155 and in this generation 
there are three component failures, which are 72, 73, and 82.
(a) The numbers on the edges are $a_{ij}$, where $i$ is the source vertex and $j$ is the destination vertex.
(b) When constructing $\mathbf{A}'$, for the two consecutive generations 72, 73, and 82
are considered to be caused by 74 since the $a_{ij}$ starting from 74 are the greatest.}
\label{a}
\end{figure}

Assume there are $m$ components in the system, 
we construct matrix $\mathbf{A}\in \mathbb{Z}^{m\times m}$ whose entry $a_{ij}$ is
the number of times that component $i$ fails
in one generation before the failure of component $j$
among all original cascades.
For each component failure in this generation
we find a component failure that most probably causes it.
Specifically, the failure of component $j$
is considered to be caused by the component failure with the greatest $a_{ij}$
among all component failures in the last generation, 
thus correcting $\mathbf{A}$ to be $\mathbf{A}' \in \mathbb{Z}^{m\times m}$, whose entry $a'_{ij}$ 
is the number of times that the failure of component $i$ causes 
the failure of component $j$.
An example is given in Fig. \ref{a}.
Then we can get the $m\times m$ interaction matrix $\mathbf{B}\in \mathbb{R}^{m\times m}$ 
whose entry $b_{ij}$ is the empirical probability
that the failure of component $i$ causes the failure of component $j$.
From the Bayes' theorem we have $b_{ij}=a'_{ij}/f_i$,
where $f_i$ is the number of times that component $i$ fails.
The $\mathbf{B}$ matrix actually determines how components 
interacts with each other.

We propose a simple model to simulate cascading failures with $\mathbf{B}$ matrix. 
Initially all components are assumed to work well 
and the cascading failure is triggered by a small fraction of component failures.
Since we would like to focus on the interaction of components 
rather than the triggering events, 
the component failures in generation 0 of an original cascade
are directly considered as generation 0 failures in the simulated cascade.
The columns of $\mathbf{B}$ corresponding to the initial failures are set zero 
since in our model once a component fails 
it will remain that way until the end of the simulation. 
Then the component failures in generation 0 
independently generate other component failures.
Specifically, if component $i$ fails in generation 0 
it will cause the failure of any other component $j$ with probability $b_{ij}$. 
Once it causes the failure of a component, 
the column of $\mathbf{B}$ corresponding to that component will be set zero. 
All component failures caused by generation 0 failures comprise generation 1. 
Generation 1 failures then generate generation 2. 
This continues until no failure is caused.

An example for electric power systems is presented 
to illustrate the proposed model. 
The original cascades are generated by the AC OPA simulation. 
We use the form of the AC OPA simulation in which 
the power system is fixed and does not evolve or upgrade. 
As other variants of the OPA model, AC OPA can also naturally produce line outages in generations; 
each iteration of the ``main loop" of the simulation produces another generation.
We choose transmission lines as components
and simulate a total of 5000 cascades 
on the IEEE 118 bus system \cite{118}, which represents 
a portion of a past American Electric Power Company transmission system.

\begin{figure}
\centering
\includegraphics[width=2.5in]{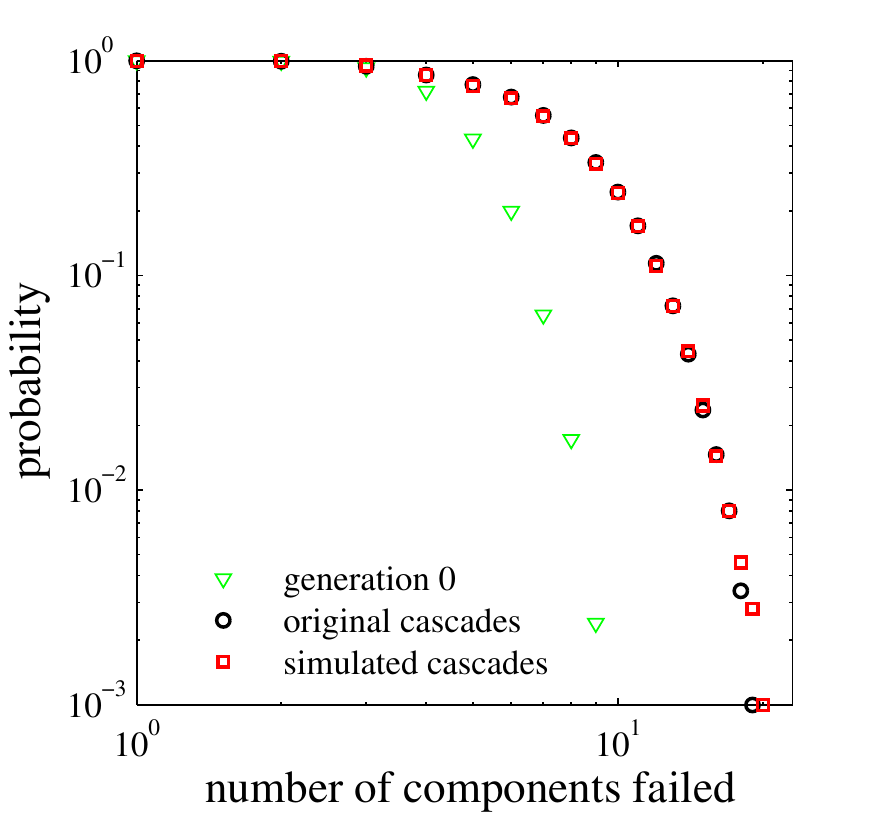}
\caption{CCD of the total number of line outages 
for original cascades (circles) and simulated cascades (rectangles). 
Downward-pointing triangles denote CCD of the generation 0 failures.}
\label{118}
\end{figure}

The complementary cumulative distributions (CCD) of the total number of line outages
for original and simulated cascades are shown in Fig.~\ref{118}.
It is seen that the two distributions match well. 
In our case $\mathbf{B}$ is a rather sparse matrix, 
a $186\times 186$ matrix with only 202 nonzero elements.
Just because of the interaction between components 
denoted by this sparse matrix, the cascading failure is 
able to propagate a lot, which is suggested by the dramatic difference 
between the generation 0 distribution and the total line outage distribution.
If the elements of $\mathbf{B}$ are all zeros and 
the components do not interact, 
all cascades will stop immediately after generation 0 failures 
and the distribution of the total line outages 
will be the same as generation 0 failures.

\begin{figure}
\centering
\includegraphics[width=2.6in]{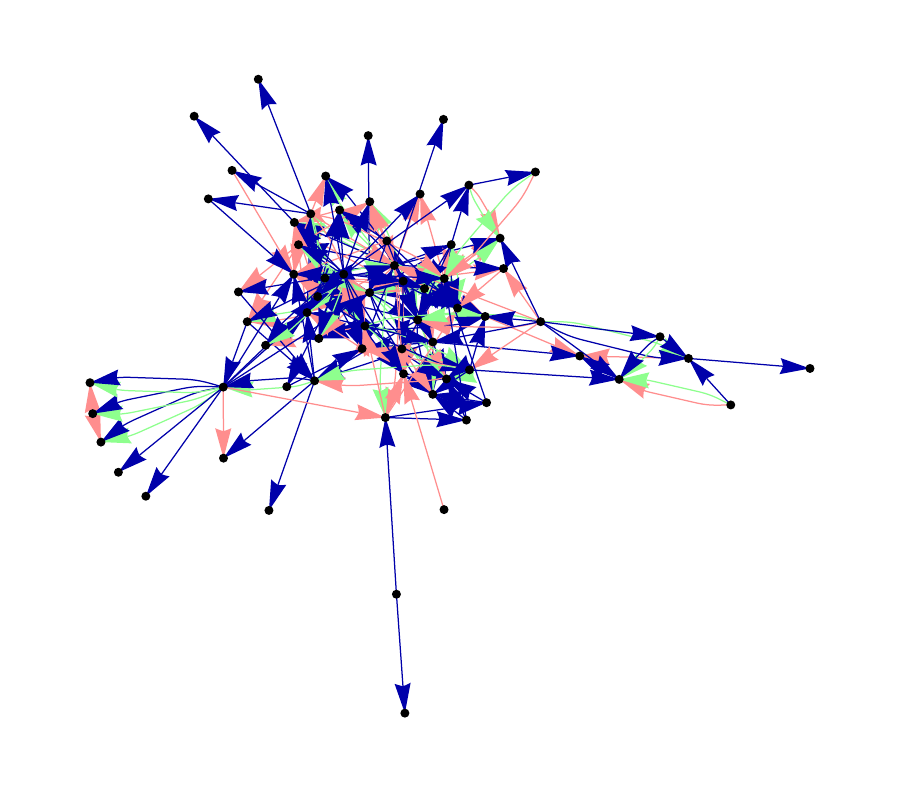}
\caption{Links for the original and simulated cascades. 
Blue links are shared by both original and simulated cascades; 
red links are only for original cascades; 
green links are only for simulated cascades; 
dots denote component failures.}
\label{links}
\end{figure}

The nonzero elements of $\mathbf{B}$ determine
how one component affects another.
They are called {\sl links}.
A link $l:\,i\rightarrow j$ 
corresponds to the nonzero element $b_{ij}$
and starts from the failure of component $i$
and ends with the failure of component $j$.
These links form a directed network,
for which the vertices are component failures
and the directed links represent that the source vertex 
causes the destination vertex 
with probability greater than 0.
Fig. \ref{links} shows the network
for both original and simulated cascades.
The shared links belong to set $L_1$
and the links only owed by the original
and simulated cascades respectively belong to set $L_2$ and $L_3$.
In our case 133 links are shared by the original
and simulated cascades. 
69 links are owned only by the original cascades 
and 45 only by the simulated cascades.

It seems that the simulated cascades
are quite different from the original cascades
since there are many different links between them.
However, it is not the truth 
if we take into account the intensities of the links.
The links are not equally important and 
can cause dramatically different consequences.
We quantify the importance of a link
$l:\,i\rightarrow j$ by an index $I_l$,
which measures the expected component failures
that a specific link $l$ can cause on the condition that
the number of times that its source vertex fails is known.

\begin{figure}
\centering
\includegraphics[width=3.2in]{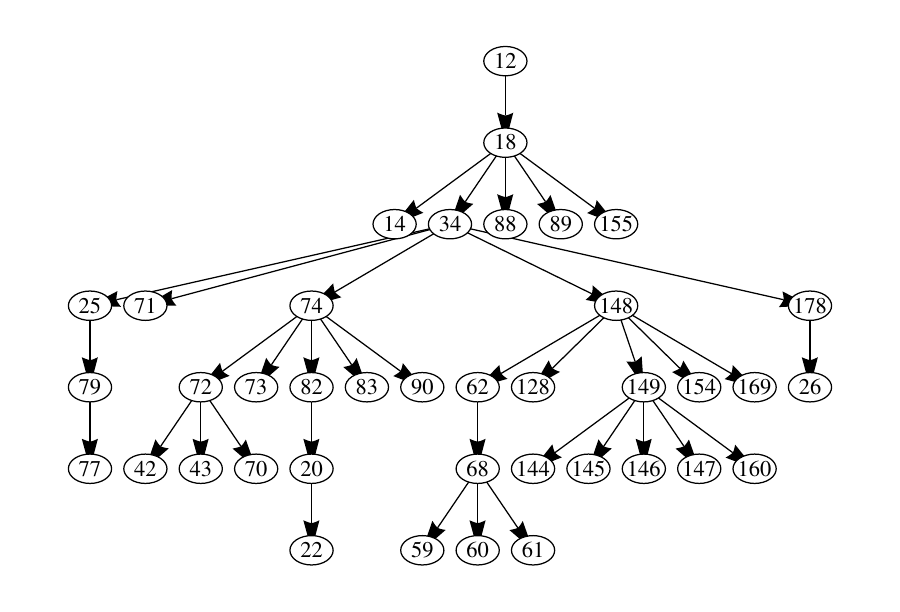}
\caption{Diagram showing how $I_l$ is calculated.}
\label{linkCalculation}
\end{figure}

Specifically, assume a component $i$ fails for $N_i$ times,
the expected failures of component $j$ will be $N_j=N_i\,b_{ij}$.
The expected failures caused by the failure of component $j$ 
is $N_j\sum_{k\in j}^{}b_{jk}$,
where $k\in j$ denotes the destination vertices starting from $j$.
We continue to calculate the expected failures
until reaching the leaf vertices.
All the expected failures are summated to be $I_l$.
In fact, $I_l=\sum_{v\in V}^{}N_v$,
where $V$ is the set of vertices for which
there exists a path starting from link $l$
and $N_v$ is the expected failures of vertex $v$.
Fig. \ref{linkCalculation} lists all vertices that 
can be influenced by link $12\rightarrow 18$.

By using $I_l$ as weights of the links
we can get a directed weighted network
corresponding to the nonzero elements of $\mathbf{B}$ matrix.
Denote the index of link $l$ for the original and simulated cascades
respectively as $I_l^\textrm{ori}$ and $I_l^\textrm{sim}$.
$\sum_{l\in (L_1 \cup L_3)}^{}I_l^\textrm{sim} / \sum_{l\in (L_1 \cup L_2)}^{}I_l^\textrm{ori}$ is 1.007,
which means that the links of the original and simulated cascades
have almost the same propagation capacity on the whole.
Besides, $\sum_{l\in L_1}^{}I_l^\textrm{ori} / \sum_{l\in (L_1 \cup L_2)}^{}I_l^\textrm{ori}$
and $\sum_{l\in L_1}^{}I_l^\textrm{sim} / \sum_{l\in (L_1 \cup L_3)}^{}I_l^\textrm{sim}$
are separately 0.975 and 0.992,
indicating that the shared links
play the major role among all links.

For the shared links
$\sum_{l\in L_1}^{}I_l^\textrm{sim}
/ \sum_{l\in L_1}^{}I_l^\textrm{ori}$ 
is 1.025. 
This suggests that the overall effects of the shared links
of the original and simulated cascades are close to each other.
However, it is still possible that the weights of the same link 
for the original and simulated cascades can be quite different.
To show if the same link is close to each other 
we define a similarity index $S$ for the shared links.

\begin{equation}
S=\sum\limits_{l\in L_1}^{}\biggr(\frac{I_l^{\textrm{sim}}
+I_l^{\textrm{ori}}}{\sum\limits_{l\in L_1}^{}(I_l^{\textrm{sim}}
+I_l^{\textrm{ori}})}\frac{I_l^{\textrm{sim}}}{I_l^{\textrm{ori}}}\biggr)
\end{equation}

When the important links of the original 
are close to their counterparts for simulated cascades, 
$S$ will be near 1.0.
In our case $S=1.071$, which means that most links, 
at least the most important links, 
have similar propagation capacity for the original and simulated cascades.

\begin{figure}
\centering
\includegraphics[width=2.5in]{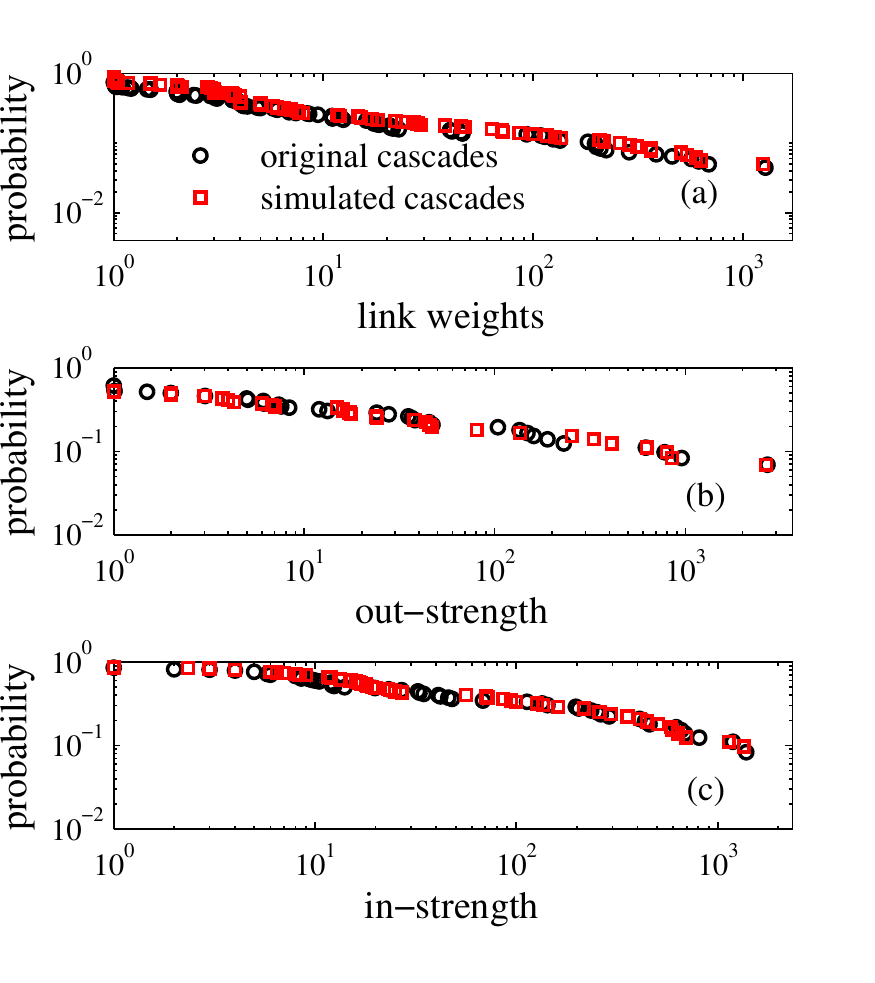}
\caption{(a) CCD of the link weights.
(b) CCD of the vertex out-strength.
(c) CCD of the vertex in-strength.}
\label{disPlot}
\end{figure}

The CCD of the link weights 
for the original and simulated cascades are shown in Fig. \ref{disPlot}(a).
The two distributions match very well. 
Both of them follow obvious power law 
and can range from 1 to more than 1000, 
suggesting that a small number of links can cause 
much greater consequences than most of the others.

The CCD of the vertex out-strength and in-strength 
for original and simulated cascades
are shown in Figs. \ref{disPlot}(b)--\ref{disPlot}(c).
Here the out-strength is defined as the summation of 
weights of the outgoing links from a vertex and 
the in-strength is the summation of weights 
of the incoming links to a vertex.
An obvious power law behavior can be seen, 
which means that most vertices (component failures)
have small consequences
while a small number of them
have much greater impact.
Another point is that the strength distributions
of the original and simulated cascades
match very well,
indicating that they share similar features
from an overall point of view.

\begin{figure}
\centering
\includegraphics[width=2.5in]{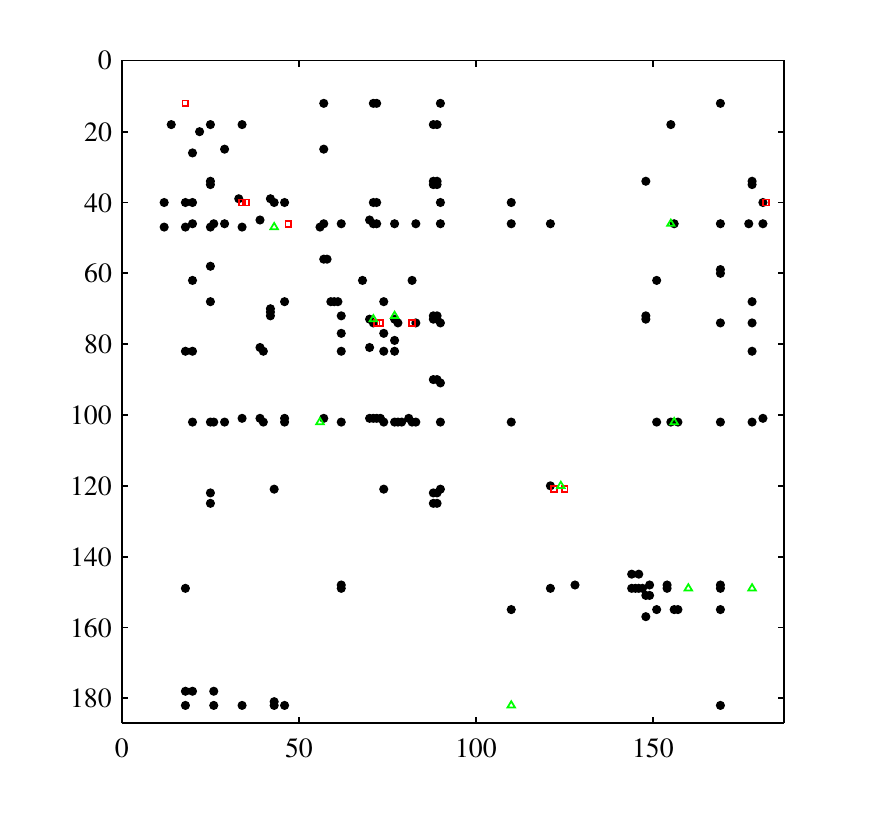}
\caption{Position of the removed links. Black dots denote nonzero elements of $\mathbf{B}$, 
among which the nonzero elements removed by intentional mitigation and random mitigation are 
denoted by red squares and green triangles.}
\label{pos}
\end{figure}

\begin{figure}
\centering
\includegraphics[width=2.5in]{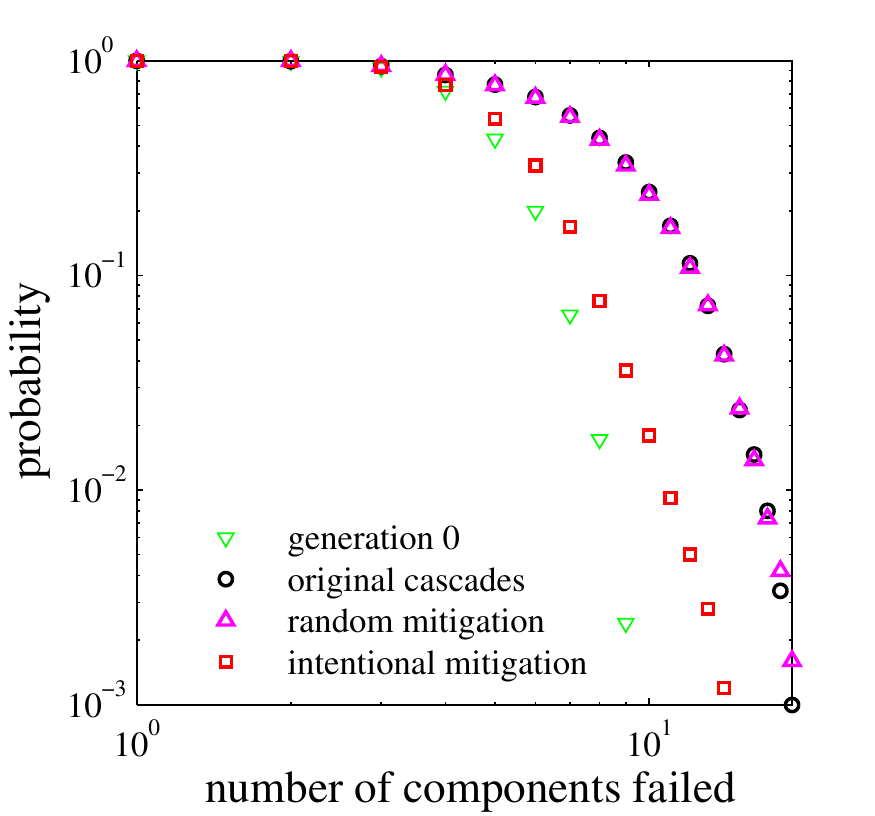}
\caption{CCD under two mitigation strategies.}
\label{mitigation}
\end{figure}

We eliminate 5\% of the links (10 links)
by setting 10 nonzero elements
in $\mathbf{B}$ matrix to be zero.
We get $\mathbf{B}_{\textrm{int}}$ by eliminating 10 links with the greatest weights 
and $\mathbf{B}_{\textrm{rand}}$ by randomly removing 10 links.
In Fig. \ref{pos} we show the position of the removed links in the interaction matrix. 
Then we separately simulate cascading failures with the proposed model by using $\mathbf{B}_{\textrm{int}}$ and $\mathbf{B}_{\textrm{rand}}$ and the two mitigation strategies are respectively called 
intentional mitigation and random mitigation. 
Fig. \ref{mitigation} shows the effects of the two mitigation strategies.
It is seen that the risk of large-scale cascading failures 
can be significantly mitigated by eliminating only a few most important links. 
By contrast, the mitigation effect is minor if we get rid of the same number of links randomly.

To conclude, we quantitatively 
determine the interaction 
of components in the system 
by calculating the probability 
that one component failure causes another.
By using this information we propose a simple model 
to simulate cascading failures.
The model is validated to be able to capture
the general properties of the original cascades
through comparison between the original and simulated cascades.
An obvious power law is found in the distributions 
of the link weights and the vertex out-strength and in-strength.
The links have significantly different propagation capacity.
A small number of links are much more crucial 
than most of the others and by eliminating them
the risk of cascading failures can be dramatically mitigated.

We are grateful for the financial support of NSFC Grant No. 50525721
and China's 863 Program Grant No. 2011AA05A118.

\end{document}